\documentclass[aps,prl,showpacs,amsmath,nofootinbib,superscriptaddress,twocolumn,preprintnumbers]{revtex4}

\usepackage[dvips]{color,graphicx}
\usepackage{amsfonts,amssymb,theorem,mathrsfs,times}
\definecolor{red  }{rgb}{1,0,0}
\definecolor{blue }{rgb}{0,0,1}
\definecolor{green}{rgb}{0,1,0}

\begin{document}
\newcommand{\M}{\mathbb{M}}
\newcommand{\R}{\mathbb{R}}
\newcommand{\HI}{\mathbb{H}}
\newcommand{\C}{\mathbb{C}}
\newcommand{\TI}{\mathbb{T}}
\newcommand{\E}{\mathbb{E}}
\def\etal{{\it et al}.} \def\e{{\rm e}} \def\de{\delta}
\def\dd{{\rm d}} \def\ds{\dd s} \def\ep{\epsilon} \def\de{\delta}
\def\goesas{\mathop{\sim}\limits} \def\al{\alpha} \def\vph{\varphi}
\def\Z#1{_{\lower2pt\hbox{$\scriptstyle#1$}}}

\newcommand{\be}{\begin{equation}}
\newcommand{\ee}{\end{equation}}
\newcommand{\bea}{\begin{eqnarray}}
\newcommand{\eea}{\end{eqnarray}}
\newcommand{\nn}{\nonumber}

\def\IR{{\hbox{{\rm I}\kern-.2em\hbox{\rm R}}}}


\renewcommand{\thefootnote}{\fnsymbol{footnote}}
\long\def\@makefntext#1{\parindent 0cm\noindent \hbox to
1em{\hss$^{\@thefnmark}$}#1}
\def\X#1{_{\lower2pt\hbox{$\scriptstyle#1$}}}


\title{Cosmology of a Friedmann-Lama\^itre-Robertson-Walker 3-brane, Late-Time Cosmic Acceleration, and the Cosmic Coincidence}

\preprint{UOC-TP 011/12, CCTP-TU/01/12}

\author{Ciaran Doolin${}^{1}$ and Ishwaree P. Neupane}

\affiliation{Department of Physics and Astronomy, University of
Canterbury, Private Bag 4800, Christchurch 8041, New Zealand}

\affiliation{Centre for Cosmology and Theoretical Physics (CCTP),
Tribhuvan University, Kathmandu 44618, Nepal}

\affiliation{CERN, Theory Department, CH-1211 Geneva 23, Switzerland}

\begin{abstract}

A late epoch cosmic acceleration may be naturally entangled with
cosmic coincidence -- the observation that at the onset of
acceleration the vacuum energy density fraction nearly coincides
with the matter density fraction. In this Letter we show that this
is indeed the case with the cosmology of a
Friedmann-Lama\^itre-Robertson-Walker (FLRW) 3-brane in a
five-dimensional anti-de Sitter spacetime. We derive the
four-dimensional effective action on a FLRW 3-brane, from which we
obtain a mass-reduction formula, namely,
$M\Z{P}^{2}=\rho\X{b}/|\Lambda\Z{5}|$, where $M\Z{P}$ is the
effective (normalized) Planck mass, $\Lambda\Z{5}$ is the
five-dimensional cosmological constant, and $\rho\Z{b}$ is the sum
of the 3-brane tension $V$ and the matter density $\rho$. Although
the range of variation in $\rho\Z{b}$ is strongly constrained, the
big bang nucleosynthesis bound on the time variation of the
effective Newton constant $G\Z{N} = (8\pi M\Z{P}^2)^{-1}$ is
satisfied when the ratio $V/\rho \gtrsim {O} (10^2)$ on
cosmological scales. The same bound leads to an effective equation
of state close to $-1$ at late epochs in accordance with
astrophysical and cosmological observations.

\end{abstract}

\maketitle

{\it Introduction}.-- The paradigm that the observable Universe is
a branelike four-dimensional hypersurface embedded in a five- and
higher-dimensional spacetime~\cite{Rubakov} is fascinating as it
provides new understanding of the feasibility of confining
standard-model fields to a D(irichlet)3-brane~\cite{Polchinski}.
This revolutionary idea, known as {\it brane-world}
proposal~\cite{RS1,RS2,Kaloper:1999,Shiromizu:1999}, is supported
by fundamental theories that attempt to reconcile general
relativity and quantum field theory,
such as string theory and $M$ theory~\cite{Horava}. 
In string theory or $M$ theory, gravity is a truly
higher-dimensional theory, becoming effectively four-dimensional
at lower energies. This behavior is seen in five-dimensional
brane-world models in which the extra spatial dimension is
strongly curved (or ``warped") due to the presence of a bulk
cosmological constant in five dimensions. Warped spacetime models
offer attractive theoretical insights into some of the significant
questions in particle physics and cosmology, such as why there
exists a large hierarchy between the 4D Planck mass and
electroweak scale~\cite{RS1} and why our late-time low-energy
world appears to be four-dimensional~\cite{IPN2010, Koyama2010}.

For viability of the brane-world scenario, the model must provide
explanations to key questions of the concurrent cosmology,
including -- (i) why the expansion rate of the Universe is
accelerating and (ii) why the density of the cosmological vacuum
energy (dark energy) is comparable to the matter density -- the
so-called cosmic coincidence problem. In this Letter, we show that
the cosmology of a Friedmann-Lama\^itre-Robertson-Walker (FLRW)
3-brane in a five-dimensional anti-de Sitter (AdS) spacetime can
address these two key questions as a single, unified cosmological
problem. Our results are based on the exact cosmological solutions
and the four-dimensional effective action obtained from
dimensional reduction of a five-dimensional bulk theory.

\medskip

{\it Model}.-- A 5D action that helps explore various features of
low-energy gravitational interactions is given by
\begin{equation}
S=  \int d^{5}x \sqrt{|g|} M\Z{5}^3 \left(R\Z{5} - 2\Lambda\Z{5}
\right)+ 2\int d^{4}x\sqrt{|h|}
\left(\mathcal{L}_{m}^{b}-V\right),\label{5d-action}\end{equation}
where $M\Z{5}$ is the fundamental 5D Planck mass,
$\mathcal{L}_{m}^{b}$ is the brane-matter Lagrangian, and $V$ is
the brane tension. The bulk cosmological constant $\Lambda\Z{5}$
has the dimension of $({\text {length}})^{-2}$, similar to that of
the Ricci scalar $R\Z{5}$.
As we are interested in
cosmological implications of a warped spacetime model, we
shall write the 5D metric {\it ansatz} in the following form %
\begin{equation}
ds^{2}=-n^{2}\left(t,y\right)dt^{2}+a^{2}(t,y)\left({dr^2\over
1-k r^2}+r^2 d\Omega^2\right) + dy^{2},\label{main-5d}
\end{equation} where $k\in\left\{ -1,0,1\right\} $ is a constant
which parametrizes the 3D spatial curvature and $d\Omega^2$ is the
metric of a 2-sphere. The equations of motion are given by
\begin{eqnarray}
G^A_B =-\Lambda\Z{5} \delta^A_B +  {\delta(y) \over M\Z{5}^3}
\mathrm{diag}
\left(-\rho\X{b},p\X{b},p\X{b},p\X{b},0\right),\label{main-5Deqs}
\end{eqnarray}
with $\rho\X{b}\equiv\rho+V$ and $p\X{b}\equiv p-V$, where $ \rho$
and $p$ are the density and the pressure of matter on a FLRW
3-brane. The parameter $h$ that appeared in Eq. (\ref{5d-action})
is the determinant of four-dimensional components of the bulk
metric, i.e., $h_{\mu\nu}(x^\mu)= g_{\mu\nu} (x^\mu, y=0)$.

\medskip
{\it Bulk Solution}.-- Using the restriction ${G}^{0}_5=0$ and
choosing the gauge $n_0\equiv  n(t, y=0) = 1$, in which case $t$
is the proper time on the brane, one finds that the warp factor
$a(t, y)$ that solves Einstein's equations in the 5D bulk and
equations on a FLRW 3-brane is given by~\cite{BDL2}
\begin{eqnarray} && a\left(t,y\right)  =  \Bigg\{
\frac{a_{0}^{2}}{2}
\left(1+\frac{\bar{\rho}_{b}^{2}}{6\Lambda\Z{5}}\right) +
\frac{3{\cal C}}{\Lambda\Z{5} a_{0}^{2}}\nn \\
&& \quad
+\Bigg[\frac{a_{0}^{2}}{2}\left(1-\frac{\bar{\rho}_{b}^{2} }{6
 \Lambda\Z{5}}\right) -\frac{3{\cal C}}{ \Lambda\Z{5} a_{0}^{2}}\Bigg]\cosh
 \left(\sqrt{{-2\Lambda\Z{5}\over 3}}\,
 y\right)\nn \\
&& \qquad -\frac{\bar{\rho}_{b}}{\sqrt{-6 \Lambda\Z{5}}}\,
a_{0}^{2}
 \sinh\left(\sqrt{{-2\Lambda\Z{5}\over 3}}\left|y\right|\right)\Bigg\} ^{{1/2}}.
 \label{nonzero-Lambda}\end{eqnarray}
where $a_0\equiv a(t, y=0)$ and
$\bar{\rho}\Z{b}=\rho\Z{b}/M\Z{5}^3$. The form of $n(t, y)$ is
obtained using $n=\dot{a}/\dot{a}_0$. 
%
The integration constant ${\cal C}$
enters into the brane analogue to the first Friedmann equation
\begin{eqnarray}
H^2 + \frac{k}{a_{0}^{2}} &=& {\Lambda\Z{5}\over 6} +
{\bar{\rho}\Z{b}^2 \over 36} + \frac{{\cal C}}{a_{0}^{4}},
\label{analog-Fried}\end{eqnarray} where $H \equiv \dot{a}_0/a_0$ is the Hubble expansion parameter.
The brane analogue to the second Friedmann
equation is
\begin{equation}
\dot{H}+ H^2 = {\Lambda\Z{5}\over 6} - {\bar{\rho}\Z{b}^2 \over
36} \left(2+ 3w\Z{b}\right) - {{\cal C}\over
a_0^4},\label{second-Friedmann}\end{equation} where
${w}\Z{b}\equiv p\Z{b}/\rho\Z{b}$ is the effective equation of
state on a FLRW 3-brane. The brane evolution equations are quite
different from Friedmann equations of standard cosmology: the
distinguishing features are (i) the appearance of the brane energy
density in a quadratic form, (ii) the dependence of $H^2$ on
$\Lambda\Z{5}$, and (iii) the appearance of the bulk radiation
term ${\cal C}/a_0^4$. If the radiation energy from bulk to brane
(or vice versa) is negligibly small, then it would be reasonable
to set ${\cal C}=0$. In the following we assume that ${\cal C}=0$
unless explicitly shown.

%
%
\medskip

{\it Cosmic acceleration}.-- With $V=$ const, the brane
energy-conservation
equation, $\dot{\rho}\Z{b} + 3 H (\rho\Z{b}+ p_b)=0$, reduces to   \\
\begin{equation}
\rho=\rho_{*}\, a_0^{-\gamma},\quad\gamma=3\left(1+{w}
\right),\label{cons-rule} \end{equation}
where $w = p/\rho$ is the EOS of matter on the 3-brane and
$\rho_*$ is a constant. With Eq.~(\ref{cons-rule}),
Eq.~(\ref{analog-Fried}) takes the following form:
\begin{equation}
\frac{\dot{a}_{0}^{2}}{a_{0}^{2}}+ {k\over a_0^2}
={\Lambda\Z{4}\over 3}+ {\bar{V} \bar{\rho}_*\over 18}
\left(a_{0}\right)^{-\gamma}+{\bar{\rho}_*^2\over 36}
\left(a_{0}\right)^{-2\gamma}, \label{mod-Fried}\end{equation}
where $\Lambda\Z{4} \equiv {\Lambda\Z{5} \over 2}+ {\bar{V}^2\over
12}$, $\bar{\rho}_*= \rho_*M^{-3}$, and $\bar{V}\equiv
V/M\Z{5}^3$. This
admits an exact solution when $k=0$, which is given by %
\begin{eqnarray}
&& \bar{\rho} = {\bar{\rho}_* \over a_{0}^{\gamma}} = { 6 H\Z{0}\over {\sinh\left(\gamma H\Z{0}
t\right) + \nu \big(\cosh\left(
\gamma H\Z{0}t\right)-1\big)}},\label{nonzero-lambda}\end{eqnarray} where $H\Z{0}\equiv
\sqrt{{\Lambda\Z{4}\over 3}}$ and $\nu\equiv {\bar{V}\over 6
H\Z{0}}$.
From this we find that the Hubble expansion parameter is given by
\begin{equation}
H={\dot{a}_0\over a_0}=  {H\Z{0} \left[\nu \sinh(\gamma H\Z{0}
t)+\cosh(\gamma H\Z{0} t)\right] \over {\nu \left(\cosh(\gamma
H\Z{0} t)-1\right)+\sinh(\gamma H\Z{0} t)}}.\end{equation} The
deceleration parameter
\begin{equation}
q\equiv - {a_0 \ddot{a}_0\over \dot{a}_0^2} =- {\dot{H}+ H^2\over
H^2} \end{equation} changes sign from positive to negative when
$\gamma H\Z{0}t \sim 1.1$ (cf. Fig. 1). This implies a transition
from decelerating to accelerating expansion. The onset time of
acceleration depends on $\nu$ but only modestly; generically, we
expect that $\nu = \bar{V}/6 H\Z{0}= \sqrt{\bar{V}^2/12
\Lambda\Z{4}} \gtrsim {O}(1)$. 
In the Randall-Sundrum (RS) limit ($\Lambda\Z{4}=0$), we find that
$a_0(t)\propto \left( 2 t+ {\gamma\nu} H\Z{0}
t^2\right)^{1/\gamma}$, which shows that the scale factor scales
as $t^{1/\gamma}$ at early epochs and as $t^{2/\gamma}$ as late
epochs. The crossover takes place when $H\Z{0} t\sim
2/(\gamma\nu)$. In the generic case with $\Lambda\Z{4}>0$, the
scale factor grows in the beginning as $t^{1/\gamma}$ (as in the
$\Lambda\Z{4}=0$ case), but at late epochs it grows almost
exponentially, $a_0(t) \propto \left[e^{\gamma H\Z{0} t}-
{2\nu/(1+\nu)}\right]^{1/\gamma}$.

%

\begin{figure}[!ht]
\centerline{\includegraphics[width=3.0in,height=1.8in]
{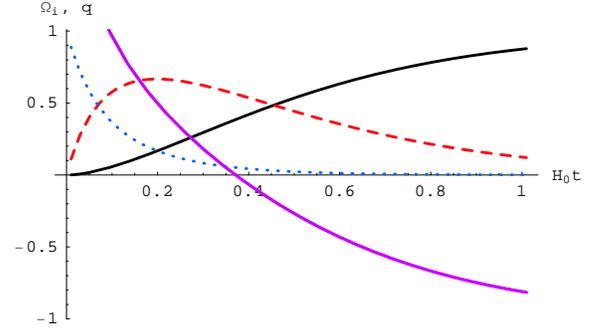}} \caption{The density fractions
$\Omega_{i}$ ($\Omega\Z{4}, \Omega\Z{\bar{\rho}}$, and
$\Omega\Z{\bar{\rho}^2}$) (solid, dashed, and dotted lines,
respectively) with $\gamma=3$ and $\nu=2$. The deceleration
parameter $q$ [solid grey (violet) line] becomes negative when
$H_0 t\gtrsim 0.37$.}
\end{figure}

\medskip
{\bf Cosmic coincidence}: Consider the Friedmann constraint
\begin{equation}
\Omega\Z{\Lambda}+ \Omega\Z{\bar{\rho}}+ \Omega\Z{{\bar\rho}^2}=1, \end{equation}
where
\begin{equation}
\Omega\Z{\Lambda}\equiv {\Lambda\Z{4}\over 3 H^2}, \quad
\Omega\Z{\bar{\rho}}\equiv {\bar{\rho}\bar{V}\over 18 H^2}, \quad
\Omega\Z{{\bar\rho}^2}\equiv {\bar{\rho}^2\over 36 H^2}.\end{equation}
As shown in Fig. 1, $\Omega\Z{\bar{\rho}^2}$ starts out as the
largest fraction around $H\Z{0} t \gtrsim 0$, but
$\Omega\Z{\bar{\rho}}$ quickly overtakes it when $H\Z{0} t\gtrsim
0.15$. Gradually, $\Omega\Z{4}$, which measures the bare vacuum
energy density fraction, surpasses these two components. Notice
that $\Omega\Z{\Lambda}+ \Omega\Z{\bar{\rho}}\simeq 1$ when
$H\Z{0} t \gtrsim 0.5$. We can see, for $\nu\simeq 2$, that
$\Omega\Z{m}\simeq 0.26$ and $\Omega\Z{\Lambda}\simeq 0.74$ when
$H_0 t\simeq 0.75$. The crossover time between the quantities
$\Omega\Z{m}\equiv \Omega\Z{\bar{\rho}}+ \Omega\Z{\bar{\rho}^2}$
and $\Omega\Z{4}$ depends modestly on $\nu $. This provides strong
theoretical evidence that dark energy may be the dominant
component of the energy density of the Universe at late epochs,
and it is consistent with results from astrophysical
observations~\cite{supernovae,WMAP7}.
Unlike some other explanations of cosmic coincidence, such as
quintessence in the form of a scalar field slowly rolling down a
potential~\cite{Steinhardt}, the explanation here of cosmic
coincidence does not require that the ratio
$\Omega\Z{m}/\Omega\Z{\Lambda}$ be set to a specific value in the
early Universe. Because of the modification of the Friedmann
equation at very high energy, namely, $H\propto \rho$, new effects
are expected in the earlier epochs and that could help to address
the challenges that the $\Lambda$CDM cosmology faces at small
(subgalaxy) scales~\cite{WMAP7}.

\medskip

{\it Effective Equation of State}.-- Eq.~(\ref{second-Friedmann})
can be written as
\begin{equation}
w\Z{b}= - {2\over 3} + {12 H\Z{0}^2\over \left(\bar{\rho} +
\bar{V}\right)^2} \left[1-\nu^2 + {q H^2 \over H\Z{0}^2}
\right].\label{wb-from-FE}\end{equation} As $H\Z{0} t\to \infty$,
$H\to H\Z{0}$, $q\to -1$, and when $\bar{V}\gg \bar{\rho}$, which
generally holds on large cosmological scales, we obtain
\begin{equation}
w\Z{b} \simeq -{2\over 3} + {1\over 3 \nu^2} \left(-\nu^2\right)
\simeq -1. \end{equation} This is consistent with the result
inferred from WMAP7 data: $w\Z{b}=-0.980\pm 0.053$ ($\Omega_k=0$)
and $w\Z{b}=-0.999^{+0.057}_{-0.056}$ ($\Omega\Z{k}\ne
0$)~\cite{WMAP7}. In the earlier epochs with $\gamma H\Z{0}
t\lesssim 1.2$, we have $w\Z{b}> -1/3$, showing that a transition
from matter to dark-energy dominance is naturally realized in the
model.

\begin{figure}[!ht]
\centerline{\includegraphics[width=3.0in,height=1.8in]
{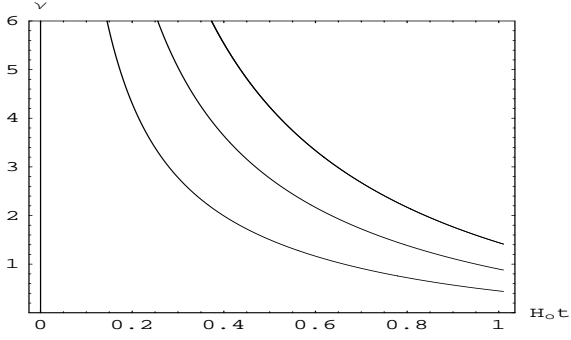}} \caption{The surfaces with ${\text
w}\Z{b}=-0.95, -0.98$, and $-0.99$ (from left to right) in the
parameter space $\{\nu, H\Z{0} t\}$.}
\end{figure}
In Fig. 2 we exhibit the parameter space for $\{\nu, H\Z{0} t\}$
with a specific value of $w\Z{b}$ at present. If any two of the
variables $\{\nu, H\Z{0} t, w\}$ are known, then the remaining one
can be calculated. Typically, if $\nu\simeq 2$ and $H_0 t\simeq
0.75$, then $w\Z{b}\simeq -0.985$. In particular, the effective
equation of state $w\Z{b}$ is given by
\begin{equation} w\Z{b} = {p\Z{b}\over \rho\Z{b}}
={p-V\over \rho +V} = {w- \zeta \over 1+ \zeta },
\label{main-def-wb}
\end{equation}
where $\zeta \equiv V/\rho$.  For brevity, suppose that the brane
is populated mostly with ordinary (baryonic) matter plus cold dark
matter, so $w \simeq 0$ ($\gamma\simeq 3$). In this case, cosmic
acceleration occurs when $\zeta > 1/2$ (or $w\Z{b}< -1/3$). This
result is consistent with the behavior of the 4D effective
potential.

\medskip
{\it Dimensionally reduced action}.-- The gravitational part of
the action (\ref{5d-action}) is
\begin{eqnarray} I &\equiv & \int
d^4{x}\, dy \sqrt{-{g}} M\Z{5}^3 \Bigg[
\frac{6}{a^{2}}\left(\frac{\dot{a}^{2}}{n^{2}}
+k-a'^{2}-a''a\right)\nn \\
&{}& + \frac{6}{an}\left(\frac{\ddot{a}}{n}
-\frac{\dot{a}\dot{n}}{n^{2}}\right) -\frac{6a'n'}{an}-\frac{2n''}{n} -
{2\Lambda\Z{5}}\Bigg],\label{effective2}
\end{eqnarray} where the prime (dot) denotes a derivative with respect to $y$ ($t$).
In order to derive from this a dimensionally reduced 4D effective
action, we may separate $a''$ and $n''$ into nondistributional
(bulk) and distributional (brane) terms
\begin{equation}
a''=\hat{a}''+\left[a'\right]\delta\left(y\right). \end{equation}
Using $n= {\dot{a}/ \dot{a}_0}$ and the
solution~(\ref{nonzero-Lambda}), the nondistributional part of the
action (\ref{effective2}) is evaluated to be-~\footnote{We
computed the integral indefinitely and then evaluated the result
at $y=0$. This approach is valid for the purpose of deriving the
time-dependent part of the 4D gravitational coupling or effective
Newton constant.}
\begin{eqnarray} && I_1 = \int d^4{x}
\sqrt{-h}\, M\Z{5}^3 {\bar{\rho}\Z{b}\over
\left(-\Lambda\X{5}\right)}\nn \\
&& \quad \times \left[ {R\Z{4}\over 2} + {\dot{\rho}\Z{b} \over 2
H \rho\Z{b}} \left( {\Lambda\Z{5}\over 2} -
{\bar{\rho}\Z{b}^2\over 12} + {{\cal C}\over a_0^4}\right) -
{\bar{\rho}\Z{b}^2 \over 9} \right],\label{distri}\end{eqnarray}
where
$R\Z{4}=6\left({\ddot{a}_{0}}/{a_{0}}+{\dot{a}_{0}^{2}}/{a_{0}^{2}}
+{k}/{a_{0}^{2}}\right)$. In the above we have employed the
background solution (\ref{nonzero-Lambda}) and integrated out the
$y$-dependent part of the 4D metric. The distributional part of
the action (\ref{effective2}) is evaluated to be
\be I_{2} = \int d^4{x} \sqrt{-h}\, {2\over 3 H}
\left(\dot{\rho}\Z{b} + 4 H \rho\Z{b}\right).
\label{backreaction}\ee The sum of $I_1$ and $I_2$ gives a
dimensionally reduced action
\begin{eqnarray}
S_{\mathrm{eff}} &=& \int\sqrt{-h} \, d^{4}x\, \left[  { M\Z{5}^3 \bar{\rho}\Z{b}\over
\left(-\Lambda\X{5}\right)} \left({R\Z{4}\over 2}
 -  \Lambda\Z{\sl eff}  \right)+  {\cal L}_m^b\right]\label{4D-final} \end{eqnarray}
with the effective potential given by
\begin{eqnarray}
\Lambda\Z{\sl eff} \equiv  {\dot{\rho}\Z{b}\over 2 H \rho\Z{b}} \left(
{5\Lambda\Z{5}\over 6} + {{\bar\rho}\Z{b}^2 \over 12}- {{\cal C}
\over a_0^4} \right)+ {{\bar\rho}\Z{b}^2 \over 9} +
{8\Lambda\X{5}\over 3} - {2\Lambda\Z{5} V\over \rho\Z{b}}\label{4d-eff-potential}.\nn\\
\end{eqnarray}
The finiteness of Newton constant is required at low-energy scale
where one ignores the effects of ordinary matter field on the
brane. In this limit, the extra dimensional volume is finite in
the same way as in canonical Randall-Sundrum models. In the
presence of matter fields, we must consider a normalized Planck
mass which generically depends on 4D coordinate time, since
$\rho_b$ is time dependent. From Eq.~(\ref{4D-final}) we read off
the normalized Planck mass
\be M_P^2 \equiv M\Z{5}^3 {\bar{\rho}\Z{b} \over (-\Lambda\Z{5})}=
{\rho\X{b}\over (-\Lambda\Z{5})}.\label{mass-red-main}\ee In the
limit that $\Lambda\Z{4} = 0$ and $V\gg \rho$,
Eq.~(\ref{mass-red-main}) reduces to the formula or identification
$8\pi G_N^{(0)} \simeq V/(6 M\Z{5}^3)$ used in~\cite{BDL2}, where
$G_N^{(0)}$ is the {\it bare} Newton's constant identified in the
low-energy limit (or when the matter density is much lower than
the brane tension). The mass reduction formula for RS flat-brane
models~\cite{RS2}, $M_P^2= M\Z{5}^3 \sqrt{-6/\Lambda}$, is
obtained as a special limit of our result, namely, $M\Z{5}^3
\Lambda\Z{5}\equiv \Lambda$, $\rho= 0$, and $\Lambda\Z{4}= 0$.

We make a remark here in regard to the scenario with $\Lambda\Z{5}=0$.
The Dvali-Gabadadze-Porrati model~\cite{DGP} corresponds
to a flat 5D bulk. In their model, it is argued that $R\Z{4}$ is generated
from loop-level coupling of brane
matter to the 4D graviton. At least at a classical level,
$R\Z{4}$ is not generated in the dimensional reduction of the
5D action if $\Lambda\Z{5}=0$, and this is exactly what we found.

\medskip
{\it AdS/FLRW-cosmology correspondence}.-- In the limit $\zeta
\equiv V/\rho\gg 1$, the 4D effective potential is approximated by
\begin{eqnarray}
\Lambda\Z{\sl eff} &=&  -{\gamma\over 2x} \left(\Lambda\Z{4}+
{{\Lambda\Z{5}}\over 3} \right)+ {4\Lambda_{4}\over 3}.
\end{eqnarray}
Note that $\Lambda\Z{4} \to {3\over 4} \Lambda\Z{\text eff}$ as
$\zeta\to \infty$. This result, which relates the bare cosmological
constant to the 4D effective potential in the limit $\rho\to 0$,
is a direct manifestation of AdS/FLRW-cosmology correspondence. In
a general case with finite $x$,
\begin{eqnarray}
\Lambda\Z{\sl eff} &=&  -{\gamma\over 2(
\zeta +1)} \left[\Lambda\Z{4}+
{{\Lambda\Z{5}}\over 3}+ {\bar{\rho}^2 \over 12} \left(2
\zeta +1\right)\right]\nn \\
&{}& \qquad ~~~~ + {\bar{\rho}^2 \over 9} \left(2\zeta +1 \right)+
{4\Lambda_{4}\over 3}+ {2 \Lambda\Z{5}\over {\zeta +1}},
\end{eqnarray}
for any value of 3D curvature constant $k$. The boundary action
(\ref{backreaction}) is crucial to correctly reproduce the RS
limit, i.e., $\Lambda\Z{\text eff}\to 0$ as $\rho \to 0$ and
$\Lambda\Z{4}\to 0$. If $\rho>0$, then $\Lambda\Z{\text eff}\ne 0$
even if $\Lambda\Z{4}=0$. This shows that the vacuum energy on the
brane or brane tension need not be directly tied to the effective
cosmological constant on a FLRW 3-brane.

\begin{figure}[!ht]
\centerline{\includegraphics[width=3.0 in,height=1.8in]
{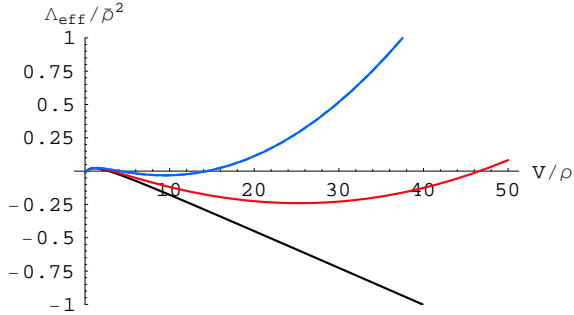}} \caption{$\Lambda\Z{\sl eff}/\bar{\rho}^2$
as a function of $\zeta =V/\rho$ with $\Lambda\Z{4}/\bar{V}^2=
0.001, 0.0004$, and $0$ (top to bottom).} \label{poten}
\end{figure}

With $\Lambda\Z{4} = 0$, there is no accelerated expansion of the
Universe, at least in a late epoch. To quantify this, take
$\gamma=4$. We then find $\Lambda\Z{\sl eff}=-\bar{\rho}^2/18
(1+\zeta )<0$, implying a decelerating Universe. If $\gamma=3$,
then $\Lambda\Z{\sl eff}>0$ in the range $ 0.177 \lesssim \zeta <
2.822 $, but in this range $w\Z{b}>-1/3$. A small deviation from
RS fine-tuning can naturally lead to accelerated expansion; the
onset time of this acceleration primarily depends on the ratio
$\Lambda\Z{4}/\bar{V}^2$. The larger deviations from RS
fine-tuning imply an earlier onset of cosmic acceleration. This
can be seen by plotting the 4D effective potential (cf. Fig. 3) or
analyzing the solution given by Eq.~(\ref{nonzero-lambda}). For
$\Lambda\Z{4}\gtrsim 0$, the model correctly predicts the
existence of a decelerating epoch which is generally required to
allow cosmic structures to form. 

{\it Constraints from big bang nucleosynthesis (BBN)}.-- In the
limit $V\gg \rho$, so $\rho\Z{b} \simeq V$,
Eq.~(\ref{mass-red-main}) is approximated as
\begin{equation}
M_P^2 \simeq {6 \over V} \,M\Z{5}^6. \end{equation} Cosmological
observations, especially BBN constraints, impose the lower limit
on $V$, namely,
\begin{equation}
V \gtrsim \left(1~{ MeV}\right)^4 \quad \Rightarrow \quad M\Z{5}
\gtrsim 10~ {\text TeV}. \end{equation}
From Eq.~(\ref{mass-red-main}) we find the time variation of the
effective Newton constant (or 4D gravitational coupling)
 \begin{equation}
{\dot{G}\Z{N}\over G\Z{N}} \simeq - {\dot{\rho} \over \rho +V} = -
{{\dot{\rho}/\rho} \over 1+\zeta }. \end{equation} From
Eq.~(\ref{nonzero-lambda}) we can see, particularly at late epochs
or when $H\Z{0} t > 0.5$, that $\dot{\rho}/\rho \to -\gamma H\Z{0}
$. The BBN bound, namely
$$
\left({\dot{G}\Z{N} \over G\Z{N}}\right)_{t_0} < 0.01 \,H\Z{0}
\sim 7.3 \times 10^{-13}\,{\text yr}^{-1},$$ which is also
comparable to current constraints from lunar-laser
ranging~\cite{JGWetal}, translates to the condition that $\zeta >
{10}^2$ (and $-1< w\Z{b} \sim  -0.99$) on large cosmological
scales.


{\it Further constraints}.-- Next we consider perturbations about
the background metric given by Eq.~(\ref{main-5d}), along with the
solution~(\ref{nonzero-Lambda}). The transverse traceless part of
graviton fluctuations $\delta g_{ij}=h_{ij}(x^\mu, y)= \sum
\varphi_m(t) f_m(y) e^{i\tilde{k}\cdot x}$ leads to a complicated
differential equation for the spatial and temporal functions,
which take remarkably simple forms at $y = 0\Z{+}$, namely,
\begin{subequations}
\begin{align}
\left(2\partial_y^2 - \bar{\rho}_b \partial_y + 2 m^2 \right) f_{i j}(0\Z{+})=0, \label{spatial}\\
\left[\partial_t^2 + 3H \partial_t + \left(m^2+{\tilde{k}^2\over a_0^2} \right)\right] \varphi_m(0\Z{+}) =0,\label{temporal}
\end{align}
\end{subequations}
where $m^2$ is a separation constant. Equation (\ref{temporal}) is
equivalent to a standard time-dependent equation for a massive
scalar field in 4D de Sitter spacetime. The masses of Kaluza-Klein
excitations are bounded by $m^2> 4H^2/9$, in which case the
amplitudes of massive KK excitations rapidly decay away from the
brane. This, along with a more stringent bound coming from
Eq.~(\ref{spatial}), implies that $H < 3 V/ (8 M\Z{5}^3)$. This is
similar to the bound coming from the background solution, namely
$\Lambda\Z{4} < \bar{V}^2/12$ or $H\Z{0} < V/(6 M\Z{5}^3)$.


{\it Conclusion}.-- Brane-world cosmology with a small deviation
from RS fine-tuning ($\Lambda\Z{4}=0$) is able to produce a
late-time cosmic acceleration. The model puts the constraints
$$ 0\lesssim {\Lambda\Z{4}\over \bar{V}^2}   \lesssim {1\over
12},\qquad \nu ={\bar{V}\over 6 H_0} \gtrsim 1.$$ The smaller the
deviation from the RS fine-tuning the larger the duration of
cosmic deceleration, prior to the late-epoch acceleration. For the
background solution (\ref{nonzero-lambda}), the brane tension is
{\it not} fine-tuned but only bounded from below. However, once
the ratio $\Lambda\Z{4}/M\Z{P}^2$ is fixed in accordance with the
observational bound $\Lambda\Z{4}/M\Z{P}^2 \sim 10^{-120}$, the
ratio $\bar{V}/ 6H\Z{0}$ also gets fixed, in which case there is a
fine-tuning between the bulk cosmological constant and brane
tension.

The method of dimensional reduction gave a simple formula,
$M_P^{2}=\rho\X{b}/|\Lambda\Z{5}|$, which relates the normalized
Planck mass $M\Z{P}$ to the matter-energy density on the brane and
the bulk cosmological constant. As $\rho\Z{b}$ is time varying,
this suggests that a mass normalized gravitational constant is
time-dependent. This is acceptable since analysis of primordial
nucleosynthesis has shown that $G_N$ can vary, although the range
of variation is strongly constrained. The BBN bound is satisfied
when the ratio $V/\rho$ is larger than ${O} (10^2)$ or when the
effective equation of state $-1< w\Z{b} \sim -0.99$.
At cosmological scales the background evolution of a FLRW 3-brane
becomes increasingly similar to $\Lambda$CDM but the model is
essentially different from $\Lambda$CDM at earlier epochs. With
precise determination of the present deceleration parameter or the
effects of a time varying equation of state, we can hope to
explore the late-time role of high-energy field theories in the
form of brane worlds and many new physical ideas.

\medskip
We wish to acknowledge useful conversations with Naresh Dadhich,
Radouane Gannouji, M. Sami, Misao Sasaki, Tetsuya Shiromizu,
Shinji Tsujikawa and David Wiltshire. I.P.N. is supported by the
Marsden Fund of the Royal Society of New Zealand (RSNZ/M1125).

\end{document}